\documentclass{emulateapj}

\def\j1517{J1517+3353}
\def\kms{km s$^{-1}$}

\slugcomment{To appear in Astrophysical Journal}

\shorttitle{SDSS J1517+3353}
\shortauthors{Rosario et al.}

\begin{document}

\title{The Jet-Driven Outflow in the Radio Galaxy SDSS J1517+3353: Implications for Double-Peaked Narrow-Line AGN}

\author{D.J. Rosario}  
\affil{University of California, Santa Cruz}
\affil{UCO/Lick, Santa Cruz, CA, 95064}
\email{rosario@ucolick.org}

\author{G.A. Shields}
\affil{University of Texas, Austin}
\affil{1 University Station, C1402, Austin, TX, 78712}
\email{shieldsga@mail.utexas.edu}

\author{G.B. Taylor \footnote{Greg Taylor is also Adjunct Astronomer at the National Radio Astronomy Observatory.}}
\affil{University of New Mexico}
\affil{MSC 07 4220, Albuquerque, NM, 87131}
\affil{gbtaylor@unm.edu}

\author{S. Salviander}
\affil{University of Texas, Austin}
\affil{1 University Station, C1402, Austin, TX, 78712}
\email{triples@astro.as.utexas.edu}

\author{K.L. Smith}
\affil{University of Texas, Austin}
\affil{1 University Station, C1402, Austin, TX, 78712}
\email{krista@mail.utexas.edu}


\begin{abstract}

We report on the study of an intriguing active galaxy that was selected as  
a potential multiple supermassive black hole merger in the early-type host 
SDSS J151709.20+335324.7 ($z=0.135$) from a complete search for
double-peaked [O III] lines from the SDSS spectroscopic QSO database. 
Ground-based SDSS imaging reveals two blue structures 
on either side of the photometric center of the host galaxy, separated 
from each other by about $5.7$ kpc. From a combination of SDSS fibre and 
Keck/HIRES long-slit spectroscopy, it is demonstrated that, in addition
to these two features, a third distinct structure surrounds the nucleus
of the host galaxy. All three structures exhibit highly-ionized line emission 
with line ratios characteristic of Seyfert II AGN. 
The analysis of spatially resolved emission line profiles from the HIRES
spectrum reveal three distinct kinematic subcomponents, one at rest
and the other two moving at -350 kms$^{-1}$ and 500 kms$^{-1}$
with respect to the systemic velocity of the host galaxy. A comparison
of imaging and spectral data confirm a strong association between 
the kinematic components and the spatial knots, which implies a 
highly disturbed and complex active region in this object.
A comparative analysis of the broadband positions, colors, kinematics 
and spectral properties of the knots in this system lead to two plausible explanations:
a.) a multiple-AGN produced due to a massive dry merger, or,
b.) a very powerful radio jet-driven outflow. 
Subsequent VLA radio imaging reveals a clear jet aligned with the emission line gas,
confirming the latter explanation. We use the broadband radio measurements to examine
the impact of the jet on the ISM of the host galaxy, and find that theÊ
energy in the radio lobes can heat a significant fraction of the gas toÊ
the virial temperature. Finally, we discuss tests that may help future surveys
distinguish between jet-driven kinematics and true black-hole binaries.  
\j1517\ is a remarkable laboratory for AGN feedback and warrants 
deeper follow-up study.

In the Appendix, we present high-resolution radio imaging of a second AGN with 
double-peaked [O III] lines, SDSS J112939.78+605742.6, which shows a sub-arcsecond radio jet. 
If the double-peaked nature of the narrow lines in radio-loud AGN 
are generally due to radio jet interactions, we suggest that extended radio
structure should be expected in most of such systems.


\end{abstract}

\keywords{galaxies: jets --- galaxies: kinematics and dynamics --- galaxies: individual (SDSS \j1517)--- lines: profiles --- galaxies: evolution}

\section{Introduction}

Active galactic nuclei are characterized by a non-stellar continuum, narrow emission lines
(FWHM $\lesssim 500~\mathrm{km~s^{-1}}$), and often broad emission lines
(FWHM $\approx 4000~\mathrm{km~s^{-1}}$).  The narrow lines generally originate in an extended
high ionization region, hundreds of parsecs in size, called the Narrow-Line Region (NLR).
In rare cases, the narrow lines from an AGN display a double peaked profile \citep{zhou04}, 
with separations between line peaks on the scale of a 100 \kms\ or greater, implying
complex kinematics of the scale of the NLR. Recent searches for such double-peaked
lines in large spectroscopic databases \citep{liu09, smith09, wang09} uncover that they 
occur in about a percent of spectroscopic AGN. 

Many explanations have been proposed for such double-peaked AGN. 
These include: unusual kinematics  in the NLR, from a bipolar jet or outflow 
\citep{whittle04, das05}, a highly anisotropic NLR geometry \citep{fws98, xu09} or 
an unresolved binary AGN, separated on NLR scales \citep[e.g.,][]{smith09}. Indeed, current
double-peaked narrow line samples may include examples of all such processes. For example,
\citet{smith09} find that radio-bright AGN are more common in their sample 
compared to a control sample of non double-peaked SDSS QSOs. A detailed study
of examples of double-peaked narrow-line AGN can help unravel ways to discriminate
between different explanations. In particular, the phenomenon of merging super-massive
black holes (SMBHs) has important implications for galaxy merger statistics and evolution, 
and narrow-line kinematics have been used to constrain merger rates at $z \lesssim 1$
\citep{comerford09}. In addition, the influence of the AGN on the evolution of its 
host is an important physical input needed to understand galaxy evolution, 
but is poorly constrained from observations. Understanding the fraction of 
double-peaked AGN that are truly merging systems or strong AGN outflows
is an important step towards defining reliable samples for studies of galaxy
merging and AGN feedback. Here, we present a study of an AGN 
with strong signatures of a double nucleus, which turns out, after careful
analysis, to be a powerful jet-driven outflow.

The intriguing nature of \j1517\ was noticed during  a systematic search of QSOs from 
SDSS Data Release 7 (DR7) to identify objects with double peaked emission-line profiles 
in  [\ion{O}{3}] and certain other emission lines \citep{smith09}. Of the approximately 
$21,000$ AGN with [\ion{O}{3}] in the SDSS spectral 
range, about 40\% had sufficient quality spectra to show double peaks of a typical 
nature.  Approximately 1\%\ of the objects had definite double-peaked narrow lines.  
Only two candidates, including \j1517, showed resolved double 
nuclei in the SDSS images. \j1517 was chosen for follow-up spectroscopy 
because its spectral line profile and appearance in the SDSS images made
it appear to be a prime candidate for harboring a double SMBH system.

An interesting point to note is that \j1517 is a Type II AGN, despite being classified
as a Type I QSO in the SDSS database. \citet{smith09}  
attribute this to a selection bias which causes double-peaked narrow line objects 
disproportionately to satisfy the SDSS requirement of an emission line wider than 
$1000$ \kms\ for classification as a QSO. For clarity, in this paper, the term 'double-peaked
narrow line AGN' refers to AGN with or without broad lines which have two well-defined
peaks in their narrow emission lines.

The organization of this paper is as follows: after introducing the observational data (\S2), 
we present basic analyses of the images and spectra (\S3-5) and a study the emission
line profiles (\S6). An ionization analysis (\S7) and presentation of new VLA data (\S8)
is followed by a discussion of the system and implications for AGN merger studies (\S9).
We assume a flat $\Lambda$CDM cosmology with H$_0 = 72$ \kms\ Mpc$^{-1}$.

\begin{figure}[t]
\figurenum{1}
\label{filterplot}
\centering
\includegraphics[width=\columnwidth]{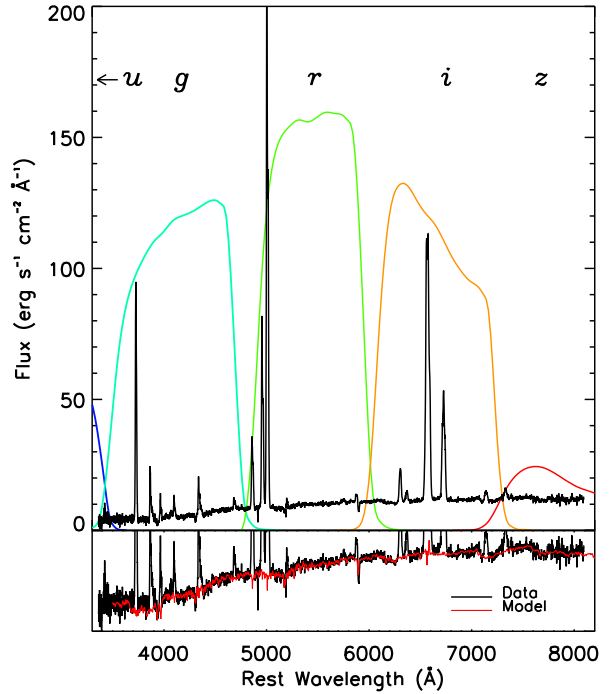}
\caption{ The SDSS spectrum of SDSS J1517+3353. Note the strong emission
lines and richness of the line spectrum. The SDSS imaging filter responses are
over-plotted. In the lower panel, a zoomed version of the spectrum highlights our
fit (in red) to the underlying continuum (in black). The quality of the fit is evident,
both in overall shape and the depths of several photospheric features.
}
\end{figure}

\section{Observations}

\subsection{SDSS and Keck Datasets} \label{sdss_keck}

The most current version of the SDSS J1517+3353 survey dataset was downloaded
from the SDSS DR7 Data Access Server (http://das.sdss.org), comprised of
images of the galaxy and its surrounding field in the five Survey filter bands 
({\it ugriz}), as well as a spectrum from the SDSS spectroscopic survey 
taken through a $3''$ fibre aperture centered on the galaxy.  
All data was processed and calibrated by the standard SDSS data
pipeline through the SDSS DAS \citep{sdssdr6}. 

A major limitation of the SDSS spectroscopic data is its lack of
resolved spatial information, which, we will show, is
necessary to disentangle the various components in this system. To gain
this added dimension, follow-up spectroscopic observations 
of the galaxy were made on the Keck I telescope, using the High-Resolution 
Echelle Spectrometer in red-optimized mode (HIRES-r).
The spectra were taken on the night of April 21 2008, in $15\arcdeg$
twilight as part of a bright object program planned for the end of the
night. The median seeing over the period of the
observations was about $0\farcs8$. Sky conditions were clear and transparent.  

A $7\farcs0$-long slit with decker C5 (slitwidth of $1\farcs15$) 
was used. The slit was placed over the centroid
of the galaxy in the telescope guide camera and aligned along a 
position angle of $122\arcdeg$ in order to cover both central peaks
in the galaxy. This PA placed the slit at an angle of $\sim 35\arcdeg$ 
from parallactic, implying that atmospheric dispersion effects, while small,
are not negligible (see \S6). 
Image grabs of the slit relative to a half arcminute field of sky around the 
galaxy were taken a few times during the observing sequence 
with the HIRES guider camera. These images were subsequently 
compared to the SDSS images to verify the alignment of the HIRES slit on the sky.

The echelle and cross-disperser grating angles were chosen to
place as many emission lines as possible on the green and red
CCD chips of the HIRES science camera. Two exposures of $600$
seconds each were taken with the same slit alignment. Quartz Halogen
and ThAr arc-lamp calibration spectra with the same instrument 
configuration as the science frames were taken at the end of
the night after sunrise. 

Spectroscopic calibration and data reduction of the raw HIRES data 
were performed using the XIDL HIRES Reduction package developed by
Jason Prochaska \footnote{Available at http://www.ucolick.org/$\sim$xavier/HIRedux/}. 
Following bias subtraction and the application of pixel-to-pixel superflats to
the science and calibration spectra, the quartz halogen lamp exposures 
were used to trace the echelle orders. Archived XIDL/HIRES ThAr arc 
templates were tweaked to match our wavelength calibration
arc-lamp exposures. These were then used to generate a wavelength solution for
each order. Finally, each curved order was extracted and rectified,
giving a set of two dimensional long slit spectra, one per exposure for
each echelle order, spanning a wavelength range from $5289 \textrm{\AA} -  
8379 \textrm{\AA}$, with small gaps in coverage redward of $6000 \textrm{\AA}$.
Cosmic rays were flagged using a sigma-clipping algorithm and the resultant
cosmic ray masks were examined visually to verify that compact subtructure in the
two-dimensional spectra were not flagged in error.

Following the observations, it was discovered that the HIRES slit had shifted
slightly between the two exposures, by less than $1\arcsec$ perpendicular
to the long axis of the slit. Since the observations were made in growing twilight, 
the guiding drift was most likely a result of the strong change
in the level of sky background between the two exposures.
The most immediate effect is a difference in the shape of the emission 
lines from the spectra taken during the two exposures. Therefore,
in our subsequent analysis, we treated each exposure independently.
Interestingly, the line profile differences between exposures 
largely affect the blue part of the lines, which provides some valuable 
insight into the spatial structure of the emission line region. Therefore, 
when appropriate, a comparative approach between the two spectral exposures
was employed in the analysis of our HIRES dataset.

\begin{figure}[t]
\figurenum{2}
\label{images}
\centering
\includegraphics[width=\columnwidth]{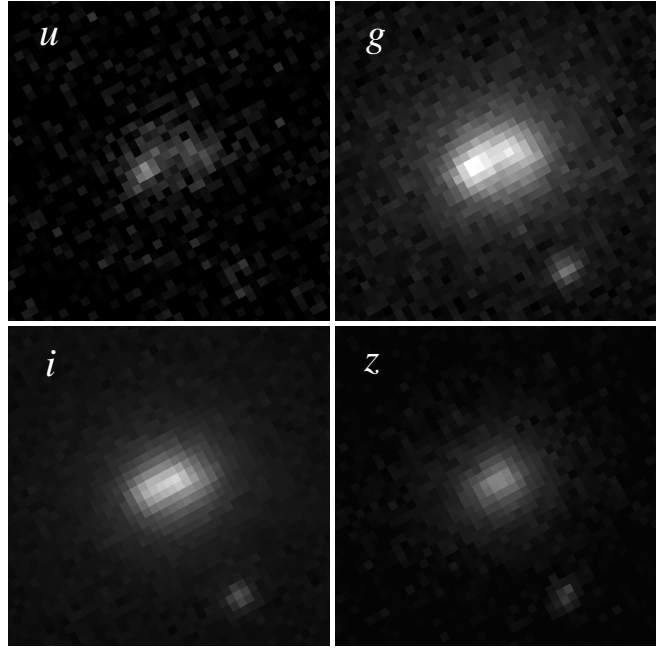}
\caption{ Panel of SDSS images of SDSS J1517+3353. The images
are $15''$ on a side; North is up and East to the left.
}
\end{figure}

\subsection{VLA dataset}

The moderately bright radio source \citep[B2 1515+34][]{colla70}
corresponding to \j1517\ was detected in the FIRST surveyÊ
\citep{becker95}, with a 1.4 GHz flux density of $106 \pm 5$ mJy, making it an ideal targetÊ
for high resolution radio imaging. Consequently, observations were taken with VLA inÊ
A-configuration on 30 October 2008 in four bands: 1.4, 5, 8, and 22 GHz.Ê
Between 5 and 10 minutes of data were obtained at each frequency. ÊTheÊ
observations were processed and reduced using AIPS in the standard fashionÊ
with flux densities tied to 3C286. ÊCLEANed images were produced with theÊ
Difmap package using natural weighting. ÊIntegrated flux densities at eachÊ
frequency band were estimated from the sum of the clean components.
Flux densities for the core components were derived in Difmap using gaussianÊ
fits to the visibility data. The images and the relation between radio andÊ
optical structures are discussed in \S8. Flux density measurements are tabulated
in Table 3. At the redshift of the galaxy ($z=0.135$) the FIRST measurement
corresponds to a monochromatic luminosity at 1.2 GHz of $5.1\times10^{31}$ erg s$^{-1}$ Hz$^{-1}$.
Comparing to the SDSS optical photometry of the host galaxy, this yields a Kellerman R radio-loudness
index of 440 \citep[defined in][]{kellermann89}, which is strictly a lower limit, since the host luminosity is dominated
by stellar light. The high value of R places \j1517\ in the category of narrow-line Radio Galaxies.

VLA images for a second radio bright AGN with double-peaked narrow lines were also taken as part of this
program. These data are presented in the Appendix.

\begin{figure}[t]
\figurenum{3}
\label{color_image}
\centering
\includegraphics[width=\columnwidth]{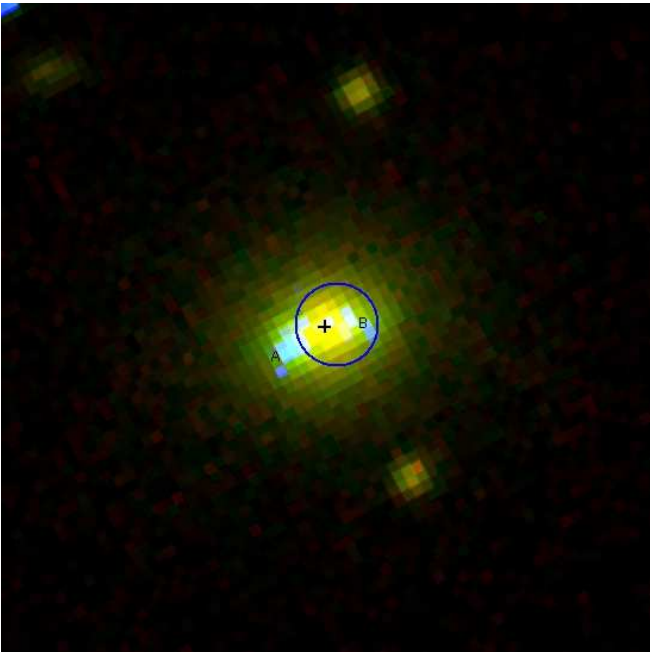}
\caption{ SDSS \emph{u'r'z'} three-color image of SDSS J1517+3353. The image
is $20''$ on a side; North is up and East to the left. A black
cross marks the center of light of the galaxy in the \emph{z'}-band. The
blue circle is the location of the circular aperture of the SDSS spectrum.
Two extended blue knots, labeled A and B, are clearly visible. A possible
third blue knot may also be seen, coincident with the center of the galaxy.
}
\end{figure}

\section{Image Analysis}

The SDSS survey images of SDSS J1517+3353 bring out the
basic structure of its active region well. 
In terms of overall S/N, the {\it g}-band image has the highest sensitivity to 
faint substructure, but careful consideration of the spectral
properties of the galaxy suggest that this is not the best image for a structural study.
In Fig. \ref{filterplot}, the bandpasses of the SDSS imaging filters are plotted
against the SDSS spectrum of SDSS J1517+3353. A number of 
strong emission lines such as [OII]$\lambda3727$, [OIII]$\lambda\lambda4959,5007$ and 
the H$\alpha+$[NII]$\lambda\lambda6584,6548$ complex may be 
identified in this spectrum. The equivalent widths
of most of these bright lines are so high that the regions of the {\it gri}  
images within a radius of $1\farcs5$ around the center of galaxy 
are strongly contaminated by line emission. The light from the galaxy in the {\it z}-band, in contrast, 
is dominated by stellar  continuum, implying that only this image actually traces the 
true shape of the host galaxy stellar distribution. This is brought out in Fig.~\ref{images}, where
we display the {\it ugiz} images of the galaxy. The first three images clearly show an elongated
central structure, mirroring the near-UV light distribution in the {\it u}-band. 
The {\it z}-band shape, however, is more circularly symmetric and uniform, 
consistent with the expected shape of an early-type host.


The SDSS {\it u}-band covers wavelengths blueward of the rest-frame Balmer break
and is dominated by Balmer and two-photon nebular continuum emission, with
relatively little light coming from stars in this part of the spectrum (see our spectral
synthesis discussion in \S4).  Consequently, the structure of the {\it u}-band image very likely traces the emission
line distribution and, by extension, the ionized gas distribution in the
galaxy. For these reasons, most measurements of gaseous or stellar substructure 
in this paper were made from the {\it u}- and {\it z}-band images respectively, despite
their lower S/N compared to the images in the other SDSS filter bands.
 
Fig. \ref{color_image} shows a {\it urz} three band color image of SDSS J1517+3353. 
While most of the galaxy is red [({\it g}-{\it i})$\sim 0.11$], the two knots on either side of the nucleus
are considerably bluer. We label them as Knots A
and B, as shown in the Figure. From the {\it u}-band
image, we estimate the separation between the approximate centroids of the blue 
knots to be $2\farcs41$, or $5.77$ kpc at the redshift of the galaxy. 
The center of the host galaxy itself is determined from the centroid of the
{\it z}-band image and is marked in Fig. \ref{images} by a cross. 
The two knots are at different distances from this center: A is $1\farcs55$ away  
while B is $0\farcs91$ away. The entire extent of the central complex of blue features
spans roughly $4\farcs84$, or about $11.6$ kpc.

It is very likely that the central several kpc of SDSS J1517+3353
contains an elongated structure that is distributed very differently
from the stars in the galaxy. While the latter is smooth, red and axisymmetric,
the former is elongated, knotty and quite blue. The spectral properties of the
galaxy's light sampled by the various SDSS filter bands hint that
the blue knots actually trace emission-line bright ionized gas rather than blue stellar
continuum from young stars, or featureless AGN light. However, this
imaging analysis is far from conclusive. For a more detailed look at the
constituents of the galaxy's circumnuclear region, we turn now to a detailed 
analysis of the SDSS spectrum.   


\section{SDSS Spectrum: Line and Continuum Measurements} \label{sdss_spec}

The location of the SDSS fibre aperture with respect to the
galaxy is shown as a blue circle in Fig. \ref{images}. The center of the 
aperture is offset slightly to the west of the actual center of the galaxy.
As a consequence, the SDSS spectrum completely samples Knot B, but does not
cover the central part of Knot A. Our analysis of SDSS line profiles in \S6
suggest that the light from some of Knot A does actually enter the fibre aperture, 
implying that the knot has an emission line distribution that is spatially resolved 
from the ground.

The SDSS spectrum (plotted in Fig.~\ref{filterplot}) covers a rest-frame wavelength range
from $3350 \textrm{\AA} - 8090 \textrm{\AA}$. The S/N of the spectrum is quite high,
with a suite of strong and weak emission lines overlayed upon 
continuum with numerous stellar photospheric features. 
Visually, the emission line spectrum is broadly characteristic of a Seyfert II AGN. 
The ratio of [\ion{O}{3}]/H$\beta$ is high, and several high excitation lines are clearly
detected, such as [\ion{Ne}{5}]$\lambda3426$, the temperature 
sensitive [\ion{O}{3}]$\lambda4363$ and \ion{He}{2}$\lambda4681$.

The richness of the SDSS spectrum, its large wavelength range and high S/N
enables us to put valuable constraints on the stellar population of the host
galaxy and the physical conditions of the line-emitting gas. Procedurally,
these analyses are inter-related. This is because the principal 
nebular diagnostics of temperature and ionization state depend on the 
accurate measurement of weak emission lines, which can be 
influenced substantially by stellar spectral features underlying the line
\citep{keel83, filip84}. Accurate spectral modeling with an emphasis on 
consistently fitting photospheric features is important and also
yields some insight into the stellar populations of the host galaxy within the 
central regions probed by the SDSS spectroscopic aperture.

We modeled the continuum using the set of galaxy template spectra 
available from the SDSS spectroscopic cross-correlation pipeline 
\footnote{Available at http://www.sdss.org/dr6/algorithms/spectemplates/index.html}. 
These empirical templates 
span a range of galaxy types, from early to late. While we are unable to constrain
such things are star-formation history and metallicity of the stellar populations
using empirical templates, this set is nevertheless ideal for the purposes of continuum
modeling, since they are based on spectra taken with the same spectrograph
as all SDSS spectroscopic data and, therefore, closely match
the wavelength range and resolution of the spectrum of SDSS J1517+3353.
Since our primary aim is the careful subtraction of the stellar continuum and not
a detailed study of the stellar populations of the host, we chose to use
the empirical template suite in the following continuum analysis.

To each galaxy template, we added a combination of 
Balmer recombination and two-photon continua,  
constrained to match the strength of the H$\beta$ line. A variable extinction correction 
in the form of a foreground screen with a Galactic extinction law was 
also applied. These modified templates were then compared to the restframe galaxy spectrum
and the extinction was varied to get the closest fit. The fitting was done by eye, since
only a small number of basic galaxy templates were finally used. 

The best fitting template was that of a
normal early-type galaxy, with an reddening of E(B-V)$\simeq 0.19$. 
In Fig. \ref{filterplot} (lower panel, we show the 
galaxy spectrum with the best-fit template overplotted in red. The fit
is remarkably good: both the general shape of the galaxy continuum and the
depth of stellar absorption features are matched extremely well over the entire
wavelength range of the spectrum. In keeping with the age of typical massive early-type
galaxies, the host's stellar population is likely to be 
predominantly old, with a median light-weighted age of several Gyrs.

The best-fit synthetic template was subtracted from the spectrum of the galaxy,
leaving a pure emission line spectrum. From this, 
we measured the strengths of all detectable emission lines. Given
the quality of the continuum subtraction, line strengths were simply derived by
adding up the flux between the wings of each emission line. Statistical errors
include uncertainties in the local continuum level and the noise variance of the spectrum.
The line flux measurements are listed in Table 1.

\begin{figure}[t]
\figurenum{4}
\label{fig4}
\centering
\includegraphics[width=\columnwidth]{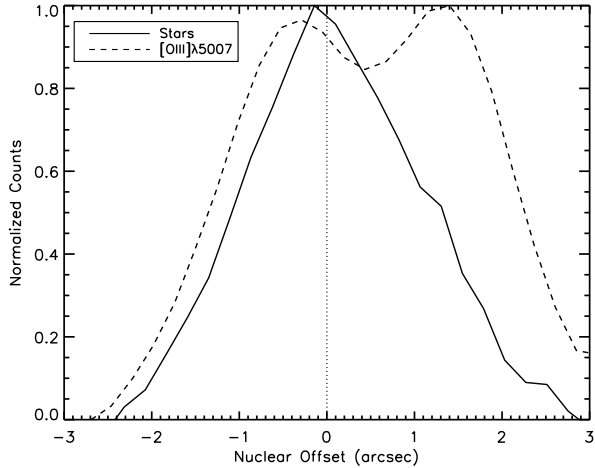}
\caption{ Spatial profiles of the [\ion{O}{3}]$\lambda5007$ (dotted lines) and
red continuum (solid line) along the HIRES slit. A correction due to the 
effects of atmospheric dispersion has been included when registering the two
profiles. We associate the median of the red continuum profile with the center
of light and the position of the nucleus of the galaxy.
}
\end{figure}

\section{HIRES spectrum: Registration}

The Keck/HIRES dataset addresses the question of whether the two distinct
velocity components in the SDSS spectrum are 
spatially distinct, as expected for a pre-merger binary AGN, and are related to the two blue knots
in the SDSS images. The high spectral resolution of HIRES
(R $\sim 37500$) imparts an added benefit -- the velocity substructure
of the line-emitting gas is exquisitely resolved;  and as we shall show, 
this brings out an additional important velocity component not apparent 
in the lower resolution SDSS data.  

However, before we undertake a full spatial and kinematic analysis, 
it is necessary to determine the exact location of the center of the galaxy 
along the long slit axis so that the structure of the emission lines
can be compared with the structures seen in the SDSS images. 
To do this, we first extracted echelle orders in the reddest parts of the
HIRES spectrum where the host galaxy stellar light is strongest and most dominant, 
while being free of significant emission line contamination. 
Since the high spectral resolution of HIRES spreads the light from the
continuum over a very large number of pixels along the dispersion 
direction, integrating over a large stretch of red spectrum is necessary
to overcome the combined noise from the sky background,
read-out and electronic shot noise.
We collapsed the red spectral orders in wavelength to obtain a profile of the
galaxy continuum light along the HIRES slit. In Fig. \ref{fig4}, this continuum
profile is plotted, along with a profile of the integrated [\ion{O}{3}]$\lambda 5007$
emission line, extracted in a similar way but integrating in wavelength only between the
wings of the line. We have applied a shift of $0\farcs7$ to the
[\ion{O}{3}] profile to correct for the difference in atmospheric refraction between 
the mean of the wavelengths used to extract the continuum profile ($8500$ \AA)
and the observed wavelength of the [\ion{O}{3}] line.
The continuum profile has a well-defined peak, corresponding to
the position of the center of the galaxy along the slit. In this way, we tie
the peak of the continuum light in the SDSS {\it z}-band image to a position
along the HIRES slit. This location serves as a reference point with which to compare
the positions of features in the HIRES and SDSS datasets. 


\begin{figure}[t]
\figurenum{5}
\label{fig5}
\centering
\includegraphics[width=0.68\columnwidth,angle=270]{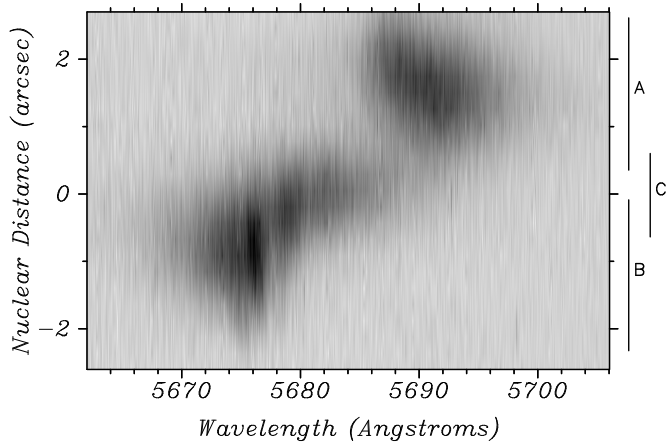}
\caption{ The [\ion{O}{3}]$\lambda5007$ line from the two-dimensional 
HIRES spectrum of SDSS J1517+3353. The cross-dispersional axis is
plotted as the angular offset from the center of light of the galaxy, along
the slit. Three kinematic features are labeled. Feature A and B correspond
to the Knots A and B from Figure 1. Feature C is contiguous with Feature B,
but has different kinematics (see Section 5.3 for details.)
}
\end{figure}

\section{Line-Profile Analysis}

\subsection{Emission Line Kinematics: Description}

The [\ion{O}{3}]$\lambda5007$ emission line from the 
HIRES spectrum of SDSS J1517+3353 is shown in Fig.~\ref{fig5}. 
The spectrograph dispersion is miniscule compared to the 
kinematic width of the line and the its shape tracks the 
kinematics of the line emitting gas within the active region. 
In general, three relatively distinct structures can be distinguished: 
two bright features separated by $14.2 \textrm{ \AA}$ or about $750$ kms$^{-1}$
with blue- and red-shifted kinematics with respect to the systemic velocity
of the galaxy, and a third component at intermediate velocities,
which spans the location of the nucleus.
By comparing the position of the features along the slit to the position of
the knots in the SDSS images of the galaxy, it is clear that the bluer 
feature corresponds to Knot B, while the redder feature corresponds to Knot A. 
We will call the third region of emission, cospatial with the nucleus,
as Feature C. There are tentative clues that this may be a distinct structure:
its kinematics appear to differ from Knot B, though both 
overlap in the sky and may possibly be contiguous in space.
Henceforth, we will use the term `Feature' to label a kinematic component
and the term `Knot' to label substructure seen in the images.
Therefore, we associated Feature A with Knot A and Feature B with Knot B.
A close examination of the color image in Fig.~\ref{images} indicates a possible
faint blue knot coincident with the nucleus, which may be associated with
Feature C.

Features A and B display fairly asymmetric line profiles. Feature A has a sharp 
blue edge and a high velocity, extended red wing. In constrast, Feature B 
exhibits a somewhat less pronounced extended blue wing. Broadly, the 
kinematic structure of Feature B is roughly antisymmetric to that of Feature A, 
and a rough mirror symmetry appears to exist between the profiles of these two features. 
However, there are some differences in detail. Feature B shows definite 
kinematic substructure, in the form of a narrow ridge of emission which extends over
the length of the Feature and defines its red edge. Feature A, on the other
hand, has a much smoother, less peaky profile. Also, Feature A shows a definite
gradient in both line asymmetry and peak velocity as a function of distance from
the nucleus. If this gradient exists along Feature B, it is far less pronounced
and not evident under the the ridge-like substructure which dominates the peak of the
line.


Visually, the kinematics of Feature C appear more symmetric, 
but its emission has to be distentangled from that
of the much brighter Feature B before a systematic study 
of its kinematics can be attempted. To do this, we adopt the
technique of gaussian decomposition of line profiles.

\subsection{Gaussian Decomposition of Emission Lines}

\begin{figure*}[t]
\figurenum{6}
\label{fig6}
\centering
\includegraphics[width=0.8\textwidth]{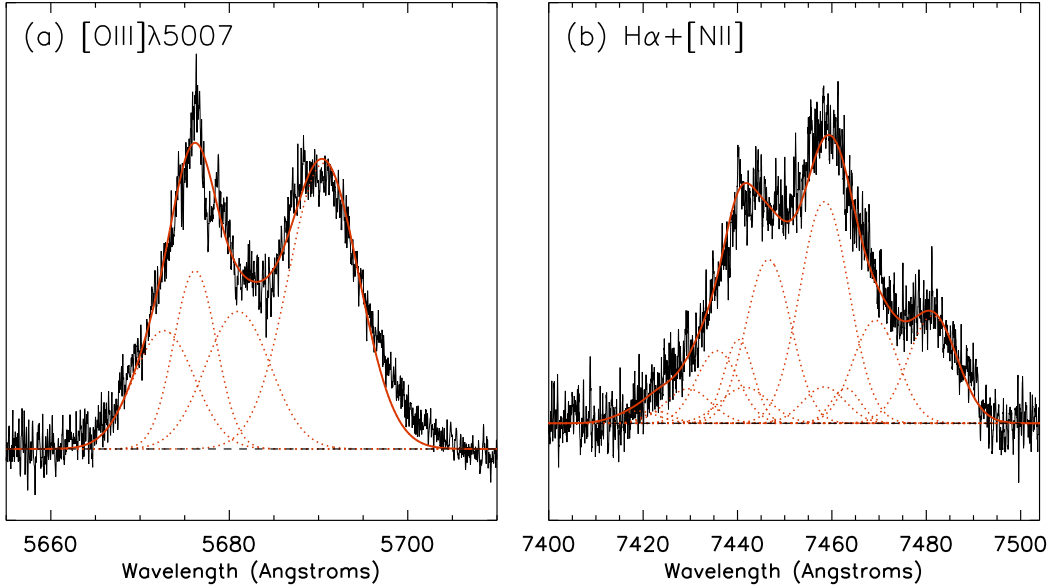}
\caption{ Gaussian subcomponent fits to emission lines in the one-dimensional
HIRES spectra of SDSS J1517+3353, collapsed along the entire length of the slit.
The observed spectrum is plotted as a black solid line, while the subcomponents
are plotted as orange dotted lines, with their sum plotted as an orange solid line.
As described in Section 5.3, four subcomponents are needed to fit the [\ion{O}{3}]
line adequately (left panel). These same velocity subcomponents, shifted appropriately, are scaled
to fit the H$\alpha$+[\ion{N}{2}] complex, as shown in the right panel. 
}
\end{figure*}

A parametrization of the kinematics of the line-emitting gas may be obtained
by fitting gaussian subcomponents to the profile of the [\ion{O}{3}]$\lambda 5007$
keeping in mind that these components may not necessarily be identified
with physically independent structures.  Not surprisingly, a single gaussian 
is insufficient to model the profile of the [\ion{O}{3}] line, either integrated 
or even at any given location along the slit. However, an inspection of the 
[\ion{O}{3}] line highlights some simplifications that serve to guide our fits. Visually, the
kinematics of Features A and B are relatively continuous along the slit, so we adopt
the following approach. First, we independently fit one to two gaussians to the line
profile for the parts of the slit that do not sample Feature C 
(nuclear distances of $1\farcs5-2\farcs5$ for Feature A and $0\farcs8-2\arcsec$ for
Feature B). Then, we adopt these same gaussians at inner nuclear radii, 
maintaining their peak wavelength and width, and scale them 
while fitting additional gaussians to Feature C. In this way, 
we can determine physically motivated gaussian fits to all 
three features and model the [\ion{O}{3}] line reasonably well (Fig.~\ref{fig6}a).

Because of its strong blue wing, Feature B requires 
at least two gaussians to describe its structure -- 
a narrow component centered at $5676.16$ \AA with a line width
of $5.58$ \AA ($294$ kms$^{-1}$), and a broad component centered at $5672.69$ \AA\ with
a line width of $8.00$ \AA\ ($423$ kms$^{-1}$).  Feature A can be adequately modeled by
a single gaussian centered at $5690.57$ \AA\ with a line width of 
$9.30$ \AA\ ($489$ kms$^{-1}$), which slightly underestimates 
the red wing, though within the tolerance of this fitting exercise.
After accounting for these three components, we are left with the emission 
from Feature C, which we fit by a single gaussian centered at $5680.97$ \AA\ with a line width of 
$8.57$ \AA\ ($452$ kms$^{-1}$). The ratio of the fluxes in the three components is measured
to be Flux(A:B:C) = 2.1 : 1.5 : 1.0. Since the various Features are larger than the HIRES slitwidth,
we do not have a good handle on slit losses and, therefore, these flux ratios are probably
not representative of the actual fluxes in the Features.

The other set of bright emission lines in the HIRES spectrum is the 
H$\alpha$,[\ion{N}{2}]$\lambda\lambda 6583,6548$ complex. These three
lines are strongly blended in the HIRES spectra due to their
large line widths. Therefore, we used the gaussian subcomponents from
the [\ion{O}{3}]$\lambda 5007$ line fits to attempt a constrained fit 
to the H$\alpha$+[\ion{N}{2}] complex, by shifting and broadening
each [\ion{O}{3}] subcomponent appropriately, and scaling all nine shifted 
gaussians in strength to fit the integrated profile of these lines, 
as shown in Fig.~\ref{fig6}b. We require that the ratio of the 
[\ion{N}{2}]$\lambda\lambda 6583,6549$ lines be
3:1, dictated by atomic physics. It may be argued that this process is
preferable as a way to fit H$\alpha$+[\ion{N}{2}], since, in the absence
of a strong dependence between gas kinematics and ionization, 
the average line profiles of all emission lines from a spatially resolved
emission line region should be very similar. From Fig.~\ref{fig6}b, it is 
clear that this approach yields very acceptable fits to these lines. 
The quality of the fit was equally good for the line profiles
extracted from both HIRES exposures, strengthening the validity
and robustness of our approach. 

The details of the [\ion{O}{3}] subcomponents and their ratios with respect
to H$\alpha$ and [\ion{N}{2}]$\lambda 6583$ are listed in Table 2. Since
our HIRES spectra are not flux-calibrated, we used the total flux
in [\ion{O}{3}]$\lambda 5007$ and H$\alpha$+[\ion{N}{2}] complex measured from
the SDSS spectrum to scale the uncorrected values from the HIRES fits,
before estimating the ratios listed in the table.


\begin{deluxetable}{lc}
\tablewidth{0.7\columnwidth}
\tablecaption{SDSS Emission Line Measurements} \label{table1}
\tablehead{ \multicolumn{1}{l}{Emission Line} & \colhead{Flux ($10^{-17}$ erg s$^{-1}$ cm$^{-2}$)} 
}
\startdata
$[NeV]\lambda3426$ & $56.2\pm8.9$ \\
$[OII]\lambda3727$ & $1390\pm69$ \\
$[NeIII]\lambda3863$ & $261\pm51$ \\
$[SII]\lambda4068$ & $51.6\pm5.2$ \\
H$\delta$ & $118\pm21$ \\
H$\gamma$ & $224\pm37$ \\
$[OIII]\lambda4363$ & $39.4\pm16.1$ \\
HeII$\lambda4686$ & $80.4\pm21.9$ \\
H$\beta$ & $574\pm29$  \\ 
$[OIII]\lambda5007$ & $4280\pm214$ \\
$[NI]\lambda5200$ & $39.7\pm5.6$ \\
HeI$\lambda5876$ & $77.4\pm6.2$ \\
$[OI]\lambda6300$ & $293\pm36$ \\
H$\alpha+[NII]\lambda\lambda6583,6548$ & $4600\pm230$ \\
$[SII]\lambda6720$ & $1410\pm80$ \\
$[ArIII]\lambda7136$ & $140\pm11$ \\
$[OII]\lambda7325$ & $171\pm21$ \\
\enddata
\end{deluxetable}

\subsection{Emission Lines in the SDSS Spectrum}

Valuable insight into the ionization and extinction properties of the gas in the Features
may be obtained by combining the results of the HIRES subcomponent fitting with the
wider suite of emission lines in the SDSS spectrum, in particular the intermediate
strength lines of H$\beta$ and the [\ion{S}{2}]$\lambda6717,6731$ doublet. 

We start with a three-component gaussian fit to the [\ion{O}{3}]$\lambda5007$ line in the
SDSS spectrum. In the case of the HIRES data, four gaussian components were needed
to completely describe the [\ion{O}{3}] line. However, the much lower spectral resolution of
the SDSS data prevents us from consistently deblending the two components that  
constitute the profile of Feature B. Therefore, we treat these two 
components as equivalent to a single one at lower resolution 
and only used three gaussians to fit the [\ion{O}{3}] profile
in the SDSS spectrum. The fit was guided by the results of the HIRES analysis. 
Given the different sky coverage of the SDSS aperture compared to the HIRES slit,  
coupled with the highly complex and nonuniform structure of the galaxy's active region,
the subcomponents that were used to fit the [\ion{O}{3}] line in the
HIRES spectrum do not exactly match the [\ion{O}{3}] line the SDSS spectrum.
However, slightly altered versions of the HIRES subcomponents, carefully broadened
to match the resolution and dispersion of the SDSS spectrum, 
provide a very reasonable fit. The properties of these subcomponents are listed
in Table 2. Note that we did not
attempt to fit the wings of the [\ion{O}{3}] line in any detail.  
The wings of emission lines from Seyfert NLRs frequently display
ionization properties that differ from the bulk of the line \citep{whittle85,veilleux91}

While H$\beta$ is detected in the HIRES dataset, it is much too weak
for a profile analysis study. Instead, we use the fit to the SDSS [\ion{O}{3}]
line, described above, as a template to fit H$\beta$, as well as few other strong
lines in the SDSS spectrum. The [\ion{O}{3}] subcomponents were shifted to the
observed wavelength of each line to be fitted, corrected for the wavelength-dependent 
differences in SDSS spectral resolution between [\ion{O}{3}] and the line, and then
scaled in strength to reproduce the profile of line. Lines fitted in 
this manner were H$\beta$, the H$\alpha$+[\ion{N}{2}] complex and the 
[\ion{S}{2}] doublet at 6720 \AA. 

To conclude, using a combination of constrained gaussian subcomponent fits
and scaled template matching, we estimate, for each spectral Feature,
kinematic parameters such as a bulk velocity and line width, as well as
the strengths of six strong emission lines in a fairly self-consistent
and controlled manner (Table 2). We are now poised to
explore the physical properties of the ionized gas in some detail.

\section{Emission Line Properties}

With these careful line measurements, we can investigate the
ionization state and physical conditions of the line-emitting gas in the Narrow-Line
Region of SDSS J1517+3353, with the final aim of discriminating between the two most likely
scenarios for the morphology and dynamics of the galaxy: a multiple black-hole merger
or a powerful jet-driven outflow.

\subsection{Physical Conditions in the NLR}

We start by evaluating some basic average properties of the line-emitting gas: its density,
temperature and dust content. The electron density of the gas can be estimated 
approximately from the ratio of the [\ion{S}{2}]$\lambda\lambda 6717,6731$ doublet. 
For both the high velocity Knots A \& B, we estimate the 
ratio $R_{\frac{6717}{6731}}$ to be $0.71\pm0.10$, which translates
to an electron density log $n_e = 3.40\pm0.24$ in cm$^{-3}$. For the nuclear Knot C,  
$R_{\frac{6717}{6731}}=0.99\pm0.14$, giving log $n_e = 2.87\pm0.25$ cm$^{-3}$. 
The densities of all three Knots are comparable, though the nuclear gas may
be slightly denser, at least in the parts of the clouds which emit most of the [\ion{S}{2}] line. 


For typical AGN Narrow Line Regions, the Balmer decrement
H$\alpha/$H$\beta$ has a value of 3.1.
Deviations from this standard value are a signature of dust extinction in, 
or along the line of sight to, the line-emitting gas. We have measured
H$\alpha$ and H$\beta$ for each individual subcomponent in \S6, 
from which, for the three Features A/C/B, we get the ratio 
of the two lines to be 5.3/4.6/3.3. This implies that Feature A is
significantly more dust reddened than Feature B, with Feature C in between.
Assuming that the dust is distributed as a foreground screen and follows a Galactic
extinction law \citep{ccm89}, we derive the visual extinction $A_V$ towards
Features A/C/B to be 1.35/0.98/0.1, after correcting by $A_V = 0.062$ for the
dust in our own galaxy towards the direction of SDSS J1517+3353.     
This may be compared to the average value of $A_V = 0.59$ implied by the spectral
synthesis analysis in Sec.~\ref{sdss_spec} 

The [\ion{O}{3}]$\lambda4363$ auroral emission line is sensitive to the electron
temperature of the ionized gas. Though weak, this line is clearly detected in the SDSS spectrum. 
We measure the ratio $R_{\frac{5007}{4363}}$ to be $87\pm36$, after applying
a reddening correction for an average extinction of $A_V = 0.9$. 
Using the standard calibration from \citet{osterbrock89} and taking a mean density
of log $n_e = 3.0$, this yields an electron
temperature estimate of $T_e = 12000^{+2300}_{-1500}$ K, somewhat
lower than a typical AGN NLR, though within the range measured in the
normal population.



Finally,  we explore whether the emission line region 
may be composed of an extended complex of H II regions from recent star formation, 
rather then AGN activity. For this, we look at line ratio combinations which 
discriminate between AGN and star-forming systems
\citep[for e.g, ][and references therein]{kewley06}. We measure
a ratio of [\ion{N}{2}$\lambda6584$]/H$\alpha$ to be 0.65/0.42/0.59
for Knots A/B/C. At these values, AGN/star-forming composite galaxies
have [\ion{O}{3}]/H$\beta$ less than 1.83/2.95/2.08, whereas we 
measure this ratio to be 7.37/7.81/6.09, placing all three knots
significantly above the transition between AGN and the star-forming
sequence. While [\ion{N}{2}]/H$\alpha$ in SDSS J1517+3353 is 
somewhat lower than the median value for Seyfert-like AGN, the high [\ion{O}{3}]/H$\beta$ 
ratio clearly places the galaxy in this general class, separate from LINERs. 
Ionization by the UV radiation from young stars is clearly not very important in 
exciting the line emission. 
This is consistent with the continuum spectral analysis of
Section 4.1, in which we found that the light of the host galaxy is dominated by
an old stellar population, with little or no evidence for blue light
from young stars, even blueward of the Balmer break. Hence,
whatever mechanism is responsible for ionizing the gas
does not contribute greatly to the visible continuum.

\begin{table*}[b]
\begin{center}
\caption{Emission Line Subcomponents}\label{table2}
\begin{tabular}{|c|ccc|ccccc|}
\hline
\hline
Comp. & \multicolumn{3}{c|}{[OIII]$\lambda 5007$} & \multicolumn{5}{c|}{Fluxes in other lines} \\
  & Velocity & FWHM & Flux & H$\beta$ & H$\alpha$ & [NII] & [SII]$\lambda 6717$ & [SII]$\lambda 6731$ \\
  \hline
B & -300 & 360 & 1250 & 160$\pm$24 & 529$\pm$106 & 221$\pm$66 & 105 & 147 \\
C & 0.0 & 450 & 1200 & 197$\pm$30 & 910$\pm$91 & 541$\pm$81 & 265 & 267 \\
A & 550 & 560 & 1460 & 198$\pm$30 & 1060$\pm$100 & 689$\pm$83 & 240 & 339 \\
\hline
\end{tabular}
\end{center}
\end{table*}

\begin{table*}
\begin{center}
\caption{VLA Radio Measurements} \label{vlatable}
\begin{tabular}{ccccc}
\hline
\hline
Band & \multicolumn{2}{c}{Beam size (asec)}  & \multicolumn{2}{c}{Flux  Density (mJy)}  \\
 & \phm{**}Major & Minor\phm{**} & \phm{**}Core & Total\phm{**} \\
\hline
1.4 GHz (L band) & \phm{**}2.75 & 1.41\phm{**} & \phm{**}92.0 & 127\phd\phm{**} \\
5 GHz \phd\phn(C band) & \phm{**}0.74 & 0.41\phm{**} & \phm{**}44.7 & 58.0\phm{**} \\
8 GHz \phd\phn(X band) & \phm{**}0.40 & 0.23\phm{**} & \phm{**}32.5 & 39.0\phm{**} \\
22 GHz \phd(K band) & \phm{**}0.14 & 0.09\phm{**} & \phm{**}17.0 & 17.0\phm{**} \\
\hline
\end{tabular}
\end{center}
\end{table*}

\subsection{Ionization of the NLR}

The richness of the emission line spectrum allows a detailed
look into the processes that ionize the emission line region of
the AGN. If the kinematics of the gas are driven primarily by a radio
jet, then shocks should be widely present throughout the NLR
and serve as a good source of ionizing photons. 
If, on the other hand, the emission line region is primarily photoionized by
merging accreting black holes, we may expect normal nuclear photoionization
to dominate the excitation of the NLR.

\begin{figure*}[t]
\figurenum{7}
\label{fig7}
\centering
\includegraphics[width=0.9\textwidth,angle=270]{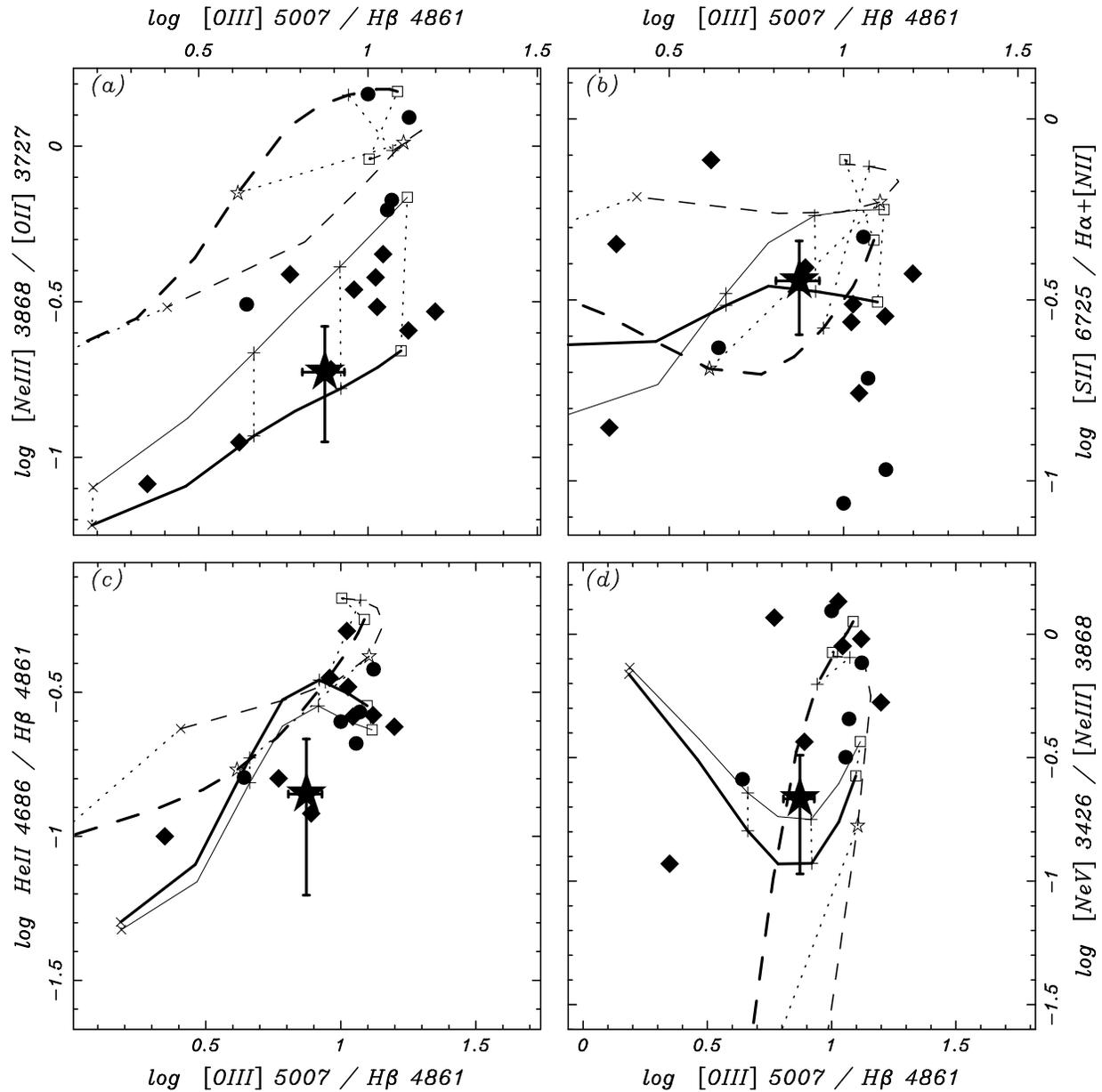}
\caption{ Emission line ratio-ratio plots comparing the ionization properties of
SDSS J1517+3353 (star point) against those of a sample of Seyferts (diamond points) and narrow-line Radio
Galaxies (circle points) from the literature \citep{koski78, cohen81}, as well as two classes of ionization
models. Shock models are plotted as solid lines, parameterized
by shock velocities which range from 200 \kms (cross) to 500 \kms (box), in steps of 100 \kms.
The thicker solid line represents magnetized shocks with a magnetic parameter of 2. 
Dusty photoionization models are plotted as dashed lines, varying in ionization parameter
marked at -4.0 (cross), -3.0 (star), -2.0 (plus),  -1.0 (box). The thicker dashed line represents a steep
ionizing spectrum ($\alpha = 2.0$). See \S7.2 for a discussion of the models and references.
}
\end{figure*}

The dynamical and physical modeling of shocks is a complex process, 
involving, among other variables, important dependences on geometry and 
magnetic fields. Shocks compress and collisionally ionize gas, 
leading to post-shock emission which produces a profuse
ionizing spectrum. These ionizing photons then propagate outward from the
post-shock region and produce an extended `precursor' region. The
emergent spectrum is therefore a combination of shock and precursor
emission. The canonical models of \citet{dopita95}, which we adopt here, 
successfully reproduce the basic features of emission line 
spectra of Seyferts and radio galaxies, as long as both shock 
and precursor emission are taken into account. 
They predict emission line ratios of most strong emission lines that are
very similar to the predictions from standard photoionization models.
Unfortunately, this means that we are not able to
use strong, well-measured lines to verify the
presence or absence of shock-ionized gas, since both shocks and
AGN photoionization models generally produce the same relative
strengths of these lines. Instead, we turn to line ratios involving
weaker lines, such as [\ion{Ne}{5}]$\lambda3426$, \ion{He}{2}$\lambda4681$ and 
[\ion{O}{1}]$\lambda6300$ which may help discriminate between the two types of models, 
as shown in Fig.~\ref{fig7}. The shock models are shown as solid lines in the Figure,
spanning shock velocities from 200 \kms\ to 500 \kms.

For comparison to the shock models, we employ a set of modern nuclear
photoionization models from \citet{groves04}, which include 
the effects of dust and a broad spectrum of cloud properties, 
and are quite successful at matching the strengths of
weak AGN lines, while generally maintaining the fit to strong lines. 
For our purposes, we fix the gas metallicity at twice solar 
(appropriate for the nuclear regions of massive early-type galaxies) and
choose a gas density of $1000$ cm$^{-3}$ to match those measured from the [\ion{S}{2}] lines.
The remaining degrees of freedom in the models are the ionization parameter $U$, defined as 
the ratio of the number densities of ionizing photons to gas at the face of a cloud,
and the shape of the ionizing continuum, assumed to be a power-law in frequency, 
with a spectral index $\alpha$. The two models tracks plotted in 
Figure \ref{fig7} as dotted lines are for $\alpha$ of $1.2$ and $2.0$: 
each track is a sequence in ionization parameter over the range $3.0 \leq U \leq 0.0$.
These tracks encapsulate most of the ionization variation in the dusty models.


Note that Fig.~\ref{fig7} employs only the integrated line strengths measured from the SDSS
spectrum, not measurements from the gaussian fitting treatment. The weak lines in the spectrum
do not have the necessary S/N for accurate subcomponent analysis. In order to place 
\j1517\ in the context of the broader AGN population, the Figure also contains data points 
from a sample of nearby Seyferts and NLRGs drawn from the literature.

In general, both sets of models match the line ratios of \j1517\ well.
The photoionization models fail badly in one case (Fig.~\ref{fig7}a), where they 
severely over-predict the strength of [Ne III]$\lambda 3869$,
considerably more than is observed in almost all the narrow line AGN on the plot. This failure has been 
noted in previous studies of Seyfert ionization \citep[e.g.][]{whittle05}.
All in all, the shock models seem to perform consistently the best in matching the measurements.
Shock velocities around $400$ km s$^{-1}$ are needed to explain the line
ratios -- at the low end, but of the same order as the measured FWHM of the kinematic subcomponents.
Based on this, we tentatively conclude that shocks are the most likely source of 
ionization in most of the emission line gas in the galaxy. 
The limitations of current NLR ionization models prevent any conclusive 
statements to be made, as photoionization models can also explain most
of the line ratios quite adequately.

In summary, the profiles of a number of emission lines from the AGN in
SDSS J1517+3353 can be characterized by a minimal set of four gaussian
subcomponents which appear to be associated with three spatially
and kinematically distinct features seen in the central $5$ kpc of the
galaxy. These features are bluer than the mean color of the host galaxy
and do not trace the galaxy's light. From our line profile analysis,
we see that the kinematics of these features are strongly disturbed, 
far in excess of any motion that may be driven by virial
processes, even in a massive early-type galaxy such as this AGN host.
Two of the features, A and B, are in overall bulk motion with respect to
the nucleus of the galaxy. Their profiles display some symmetries, but the 
relative distance of these features from the nucleus are quite different, with 
Feature B being significantly closer to the center of the galaxy than Feature A.
Their gas excitation properties are similar and produce a highly ionized
Seyfert-like spectrum, though Feature A is associated with much greater
dust extinction, either instrinsic or along the line of sight, than Feature B.
The third feature, C, is coincident with the galaxy's center of light
and probably surrounds the true nucleus of the galaxy. Unlike Features A
and B, Feature C has a kinematic profile which is quite symmetric. Its excitation
properties are a bit different from the two off-nuclear features, with stronger
lines from low ionization species.

\section{VLA maps}

\begin{figure*}[t]
\figurenum{8}
\label{vlamaps}
\centering
\includegraphics[width=\textwidth]{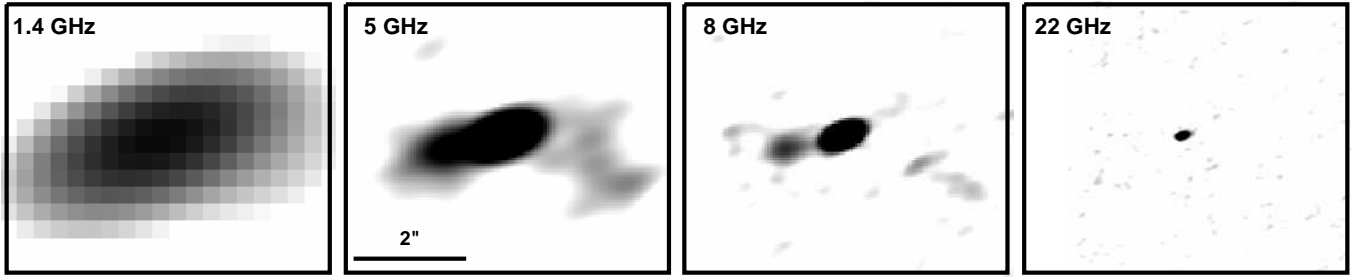}
\caption{ A montage of VLA radio images of \j1517\ at four different wavebands. The effective semi-major
beam-size of these maps drops from $2\farcs75$ at 1.4 GHz to $0\farcs15$ at 22 GHz.  The contrast
of the jets with respect to the core drops strongly in the high frequency maps - a consequence of the steeper
spectrum of the radio jets.
}
\end{figure*}

\begin{figure}[t]
\figurenum{9}
\label{vlaoverlay}
\centering
\includegraphics[width=\columnwidth]{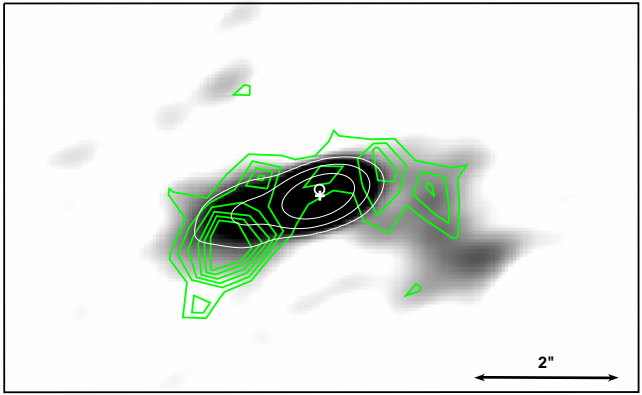}
\caption{ Contours of the SDSS \emph{u}-band image (in green) plotted over the 5 GHz radio map. 
Thin white contours bring out the gradient in the core of the radio source. North
is up and East is to the left. A small cross and a circle mark the centers, respectively, of the radio
core and the host galaxy (from the SDSS \emph{z}-band image).
Note the strong alignment between ionized gas, traced by the \emph{u}-band, and radio structures, 
including the bend in the western jet.
}
\end{figure}

We present four VLA maps of \j1517\ in Fig.~\ref{vlamaps} at each of the four bands: 1.4, 5, 8 and 22 GHz. 
The maps clearly show that the radio source is elongated, with a moderately bright (32.5 mJy at 8 GHz)
unresolved flat-spectrum core, and a steep-spectrum bipolar jet which disappear at the highest frequencies 
(22 GHz).  The astrometric position of the radio core is 15:17:9.24228, +33:53:24.5663 (J2000), accurate
to 25 mas. The eastern jet is brighter than the western jet and appears more collimated.
Of the four VLA images, the 5 GHz map provides the best combination of resolution and sensitivity 
to extended structure and we adopt this map to compare the radio structure to the optical structures in the NLR. 
In Fig.~\ref{vlaoverlay}, contours of the 5 GHz map are overlayed on the SDSS u-band image, with native
astrometric information used to align the images. The center of the VLA core lies $0\farcs17$ (0.42 
SDSS image pixels) south of the location of the nucleus in the z-band image. Given the restrictions 
imposed by the seeing limited resolution and the S/N of the SDSS images, we hesitate to attribute 
this offset to any real difference between the nuclear positions of the galaxy and the radio jet. 
Instead, we assume that our error in absolute registration between the SDSS and VLA 
imaging datasets is of the order of $0\farcs2$. The principal component
of this error is the uncertainty of the nuclear position measured from the SDSS z-band image.

Despite the registration uncertainty, it is immediately clear that there is a 
close similarity in structure between the radio jet and the emission
line gas. The western jet shows a strong bend of about $75^{\circ}$ at $1''$ from the core. This bend is traced
by the emission-line gas, such that Knot B is cospatial with the location of the jet bend. Knot A appears to
border the southern edge of the eastern jet. Based on the extreme kinematics, ionization conditions and physical
proximity of the jets and the emission line features, we conclude that the radio source is strongly interacting with
gas on kpc scales in the galaxy. This gas is being accelerated and possibly ionized by strong shocks driven by
this interaction. Similar structures are found in other radio galaxies  at comparable and higher redshifts
\citep{tadhunter89,solorzano01} and a relationship between extended radio jets and the disturbance of 
AGN emission line regions has been known for decades \citep{heckman81, whittle92b}.

As noted in \S6.1, Feature B has a narrow ridgeline in its [\ion{O}{3}] profile, which appears around
the nucleus and extends to about $1\farcs5$ (Fig.~4). Feature A, while being broadly antisymmetric
with Feature B, does not appear to show any such ridgeline. This may be related to a difference
in the mode of jet-gas interaction between the western and eastern jets. The western jet 
bends sharply as it emerges from the nucleus. If the bend is produced by a strong interaction 
between the initially unbent jet and a dense gas cloud near the nucleus, it will be a site
of high pressures and strong jet-gas interaction signatures. In such a scenario, the
narrow ridge of emission may be a localized area of coherently accelerated ionized gas
within the broader outflow. Such a system is known already in the radio Seyfert Mkn 78 \citep{whittle04}.
In this galaxy, a bend in the radio jet is linked with a region of strong line splitting, which
extends for hundreds of parsecs from the location of the bend. A very similar process
may be occurring at the bend in \j1517. The eastern jet, on the other hand, does not
bend and the associated Feature A is smooth, with no conspicuous velocity substructure.
We speculate that the difference in radio brightness and the level of collimation between
the two jets may also be a consequence of a strong localized interaction in the western jet.

\section{Discussion}

\subsection{The Nature of SDSS J1517+3353}

Despite its initial selection as a candidate for a multiple merging black hole system, several 
lines of evidence confirm that the double-peaked lines of \j1517\ are due to a strong 
emission line outflow produced by a jet-ISM interaction. 
The radio jet and emission line gas are co-aligned and appear to
trace each other quite closely. This is characteristic of Radio Galaxies and Seyferts
that show jet-ISM interaction signatures \citep[][and references therein]{whittle04}.
The profiles of the kinematically distinct components appear to show a particular mirror symmetry
(\S5.3), consistent with gas accelerated by a bipolar flow or jet.
In addition, the emission line spectrum is consistent with ionization by strong shocks with a 
mean velocity of $400$ \kms, consistent with the line widths of the emission line sub-components.
Finally, the host galaxy does not appear to display any obvious signatures of recent merger activity.
The isophotes are quite regular and the galaxy colors outside the emission line region are quite red.
The stellar continuum appears to be that of an early-type galaxy with some extinction in the central
regions, presumably due to dust near the nucleus. There is no sign of the substantial 
recent star-formation which may indicate a gas-rich merger of the sort that can host merging accreting SMBHs.
 
\subsection{Analysis of the Interaction}

Through a combination of spectral line decomposition and high resolution radio
imaging, it is possible to now associate the eastern and western radio jets with
specific emission line knots. In this way, a comparison can be made between
the dynamics of the emission line gas and the energy content of the radio source.

We begin by estimating the mass and kinetic energies of the ionized gas. The 
extinction corrected Balmer line luminosity is a direct measure of the total amount of 
ionized gas:

\begin{equation}
M_{em} = 171\, F_{H\beta}\,d^2 \, n_{em}^{-1} \qquad \textrm{g}
\end{equation}

\noindent where $d$ is the luminosity distance of the galaxy in cm 
and $F_{H\beta}$ is the H$\beta$ flux. 
Using the measurements in Table 2, the masses of Feature A/B are 
both estimated to be around $10^{5.9}$ solar masses.
Taking the peak velocity and FWHM as a measure of the motion in the 
gas, we calculate bulk and turbulent kinetic energies. These are combined
to yield total kinetic energies in Features A and B of $10^{54.4}$ and 
$10^{54.1}$ ergs respectively.

The internal energy of the ionized gas is related to its density and temperature, and can be
parametrized by the mass of ionized gas:

\begin{equation}
E_{th} = 1.25\times 10^{8}\; M_{em}\; \mu^{-1} \; T_{e}  \qquad \textrm{ergs}
\end{equation}

\noindent where $\mu \sim 0.6$ is the mean molecular weight of the gas. For the
Features A/B, we get $E_{th} \sim 10^{51.5}$ for both. This is 2.5-3 orders of magnitude
lower than the kinetic energy of the gas. If we assume that the emission line luminosity 
of the gas is $L_{em} \approx 10 \times L_{[O III]} \sim 10^{43.5}$ erg/s, we derive a timescale
at which this internal energy is radiated away $\tau_{th} \approx E_{th} / L_{em} = 3$ years.
It is clear that the thermal energy of the ionized gas has to replenished, either
through jet-driven shocks or photoionization from the central AGN.

The ionized gas energies may be compared to that of the radio source.
A listing of the measured flux densities
in the four VLA bands in given in Table 3. The mean radio spectral index of the extended
emission between 5 and 8 GHz is $\alpha_{R} = -1.5$. Assuming that this synchrotron spectrum
extends between 0.01 GHz and 100 GHz, the total radio luminosity ($L_R$) of the jets can be estimated
by normalizing the spectrum to the flux of the jets at 5 GHz, which gives $L_{R} = 10^{42}$ erg s$^{-1}$.
From the 5 GHz image, we estimate the size of the eastern jet to be about $1\farcs2 \times 1\farcs1$, which,
if we assume a cylindrical jet geometry, gives a radio-emitting volume of $V_{jet} \approx 15.2$ kpc$^{3}$.
Combining this with equipartition energy arguments of \citet{miley81}, and assuming that the
radio emitting plasma has a filling factor $\sim 1$, mean relativisitic pressures, 
magnetic fields and energies are estimated to be $P_{me} = 10^{-9.4}$ dyne cm$^{-2}$, 
$B_{me} = 125$ $\mu$G and $E_{me} = 10^{56.7}$ ergs. The synchrotron 
lifetime of the radio source $\tau_{R} \approx E_{rel}/L_{R} \sim 16$ million years, comparable
or longer than a typical lifetime for Seyfert-like activity, and much smaller than the timescale
on which the radio-source imparts energy to the emission line gas through shocks and pressure-driven 
acceleration (see below). Synchrotron emission is probably not a major loss mechanism for the 
energy in the radio plasma. In addition, the energy content of the radio source, 
assuming equipartition, is more than two orders of magnitude higher than the kinetic
energy content of the ionized gas. Only a small fraction of the energy in the lobes needs to
go into accelerating the ISM of the galaxy, through radio source expansion or ram pressure-driven
shocks, to fully account for the dynamics of the emission line material. 
We envision a scenario where the jet propagates rapidly to kpc scales, at velocities much 
larger than the ionized gas, sweeping up the ISM of the host galaxy and shocking
it as it expands. The denser component of the shocked ISM in direct contact with the
jet is visible as the NLR. There may be a potentially large fraction of shocked gas
which cannot cool efficiently by line emission and may remain invisible 
\citep[e.g.][]{mellema02}. Therefore, the amount of energy from the jet that is lost to
line emission is highly unconstrained.

The pressure of the relativistic plasma, acting over 
the area of the jet, is the principal source  of momentum for the ionized gas. 
A rough interaction timescale can be derived by equating $P_{me}$ to the measured bulk momentum in the gas:

\begin{equation}
\tau = \frac{M_{em}\; v_{em}}{P_{me}\; A_{jet}}   \qquad \textrm{seconds}
\end{equation}

\noindent which yields $\tau \sim 10^{5}$ years, for $A_{jet} = 5.3$ kpc$^2$ and an average
bulk flow velocity of $v_{em} \approx 400$ \kms. The timescale for the acceleration of
the gas could be significantly smaller if the major part of the acceleration 
happened at sub-kpc radii in the galaxy, where $P_{me}$ is expected to be higher. 
However, the gas would then have to coast out to where it is currently observed on kpc scales
at or below its measured bulk velocity of $v_{em}$, adding more than 
$10^{6}$ years to the overall timescale for the
interaction. The timescale would also be longer if the accelerating force was smaller
than determined by $P_{me}$, or if the accelerated gas contained a substantial neutral 
component not detected in emission lines, as implied by some 
studies \citep[e.g.][]{morganti05}. Therefore, $\tau$ is 
probably a lower limit to the actual time that the jet interaction has been 
happening in this system.


An upper limit to the energy flux $F_E$ of the jet then follows:

\begin{equation}
F_{E} \quad = \quad E_{me}/\tau \quad = \quad 1.65 \times 10^{44} \: \textrm{erg s}^{-1}
\end{equation}

This flux is about an order of magnitude larger than the emission line luminosity and two orders
of magnitude greater than the radio luminosity. If the timescale of the interaction is indeed
as short as $\tau$, we infer that only a small fraction of the energy output
of the jet is lost to shock-excited radiation or other radiative mechanisms.   

How does the energy in the lobes compare to the binding energy of the host galaxy? We estimate
a velocity dispersion of the galaxy from the NaD absorption line, which is strong and well-defined
in the SDSS spectrum. A gaussian fit to the line yields a FWHM of 14.76 \AA\, which gives a 
velocity dispersion $\sigma_* = 274$ \kms, after a small correction for the SDSS spectrograph
resolution.  The amount of mass that can be unbound by the energy content of the radio source (i.e, $E_{me}$)
is given by:

\begin{equation}
M_{bind}  \; \approx \; \frac{E_{me}}{\sigma_{*}^{2}}  
\end{equation}

\noindent where we have taken the gravitational binding energy of a unit mass of gas in the 
central regions of the galaxy to be approximately $\sigma_{*}^2$ ergs gm$^{-1}$, assuming that
the underlying mass density traces a singular isothermal sphere.

Combining our estimates of $E_{me}$ and $\sigma_{*}$ gives $M_{bind} \sim 10^{8.5}$ M$_{\odot}$
which should be accurate to $0.5$ dex, given our uncertainties in the mass distribution of the
host galaxy and the velocity dispersion. The energy in the lobes can easily unbind 
the total mass in ionized gas ($\sim 10^{6} \textrm{ M}_{\odot}$). If we assume a gas fraction in
the galaxy of $\sim 10^{-2}$ and taking its stellar mass to be $10^{11} \textrm{ M}_{\odot}$,
the gas content of the galaxy, in both hot and cold ISM phases, is estimated to be
$10^{9} \textrm{ M}_{\odot}$. The energy in the radio lobes can unbind a good fraction of all the gas in the galaxy
and is certainly capable of heating most of this gas to its virial temperature, as long as it is
effectively coupled with the ISM. The synchrotron lifetime of the lobes sets a simple upper
limit to the efficiency of thermalization of the lobe energy. If the radio source is to serve
as a source of feedback to the hot atmosphere of the host galaxy, then the energy
in the non-thermal population of the lobes has to be exchanged with the surrounding
gas within about 16 million years, or else most of it will be lost to radiation, which
couples inefficiently with the hot gas.




\subsection{Strong AGN Outflows: Implications for Binary SMBH Searches}

\j1517\ serves as a case study of a powerful AGN outflow that effectively masquerades as
a candidate for merging accreting SMBHs. Only through spatially-resolved spectroscopy
and radio imaging is its true nature revealed. In the presence of a jet or strong
nuclear outflow, it is much more likely that complex line kinematics on kpc scales are
the result of gas interacting with the outflow rather than from inspiralling black holes.
\citet{smith09} find that their double-peaked AGN sample includes more FIRST detected 
radio bright AGN than the parent QSO population, implying
that a non-negligible fraction of the split lines probably come from jet interactions. 
\citet{whittle92b} points out that clear non-gravitational kinematics are seen even in radio-quiet Seyferts 
with 20 cm luminosities $ > 10^{22.5}$ W Hz$^{-1}$.  We suggest that a radio prior be
applied to double-peaked samples to identify such outflows. A radio galaxy like \j1517\ can be detected
to $z\approx1$ in wide-area 20 cm mJy-level surveys like FIRST and NVSS,
while radio-quiet sources above the threshold from \citet{whittle92b} are detectable with 100 $\mu$Jy 
pointed radio observations out to $z \approx 0.3$ and with wide-area surveys to $z \approx 0.1$. 
Modestly deep follow-up of double-peaked narrow-line samples with existing radio facilities to
separate outflows from true black hole mergers.

About 10\% of  local optically-selected Seyfert galaxies have total radio luminosities
greater than $10^{22}$ W Hz$^{-1}$ \citep{debruyn78, rush96, ho01}, significantly higher
than the fraction of SDSS AGN which show double-peaked profiles. If only one in ten AGN with
bright nuclear radio sources have extended jets, outflows can conceivably account for most of, if
not the entire, double-peaked population. Clearly, stronger constraints on the radio content
of these samples is needed before they can be used to study merger rates and SMBH
evolution.

On the other hand, we have demonstrated that double-peaked emission can indicate the existence of powerful, galaxy-wide outflows, where massive AGN feedback is caught in the act. 
These samples may dentify the best laboratories with which to directly 
tackle the details of feedback physics, such as the importance of momentum-driven 
winds, the effects of anisotropic obscuration, and the efficiency
of coupling between the output of the AGN and material in the host galaxy.

\acknowledgements

D.R. acknowledges the support of NSF grants AST-0507483
and AST-0808133. G.S. gratefully acknowledges support from the Jane and Roland Blumberg 
Centennial Professorship in Astronomy at the University of Texas at
Austin. Funding for the Sloan Digital Sky Survey (SDSS) has been provided by the Alfred P.
Sloan Foundation, the Participating Institutions, the National Aeronautics and Space Administration,
the National Science Foundation, the U.S. Department of Energy, the Japanese
Monbukagakusho, and the Max Planck Society. The SDSSWeb site is http://www.sdss.org/.
The SDSS is managed by the Astrophysical Research Consortium (ARC) for the Participating
Institutions. The National Radio Astronomy Observatory is operated by
Associated Universities, Inc., under cooperative agreement with the
National Science Foundation.

\appendix

\section{VLA Imaging of SDSS J112939.78+605742.6}

\begin{figure*}[t]
\figurenum{10}
\label{appendixfig}
\centering
\includegraphics[width=\columnwidth]{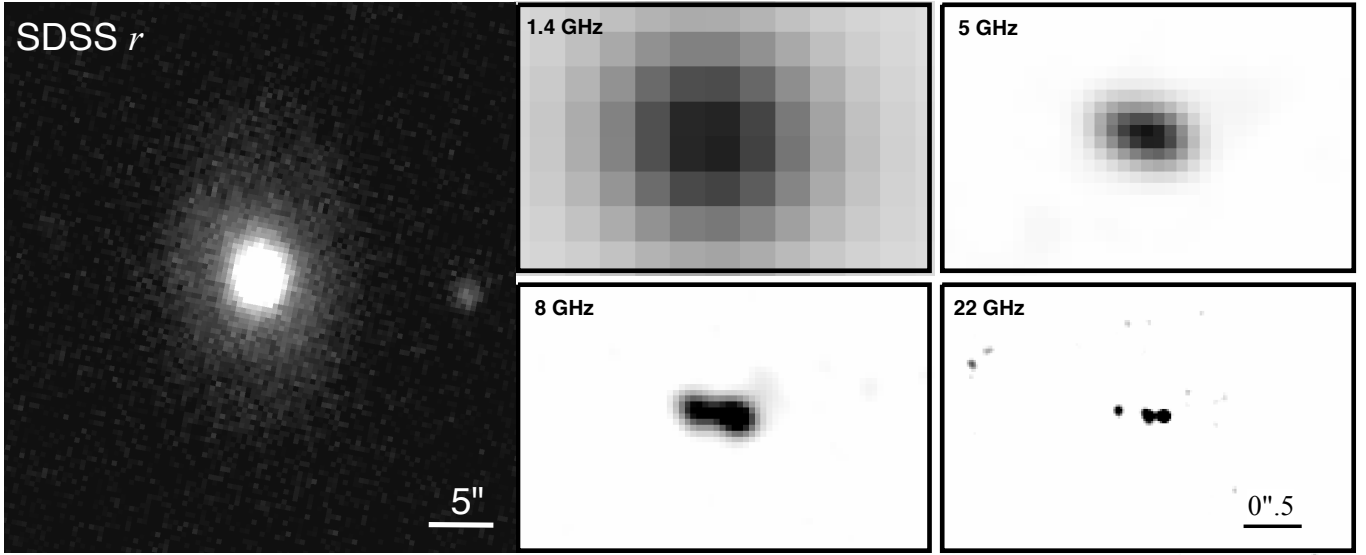}
\caption{ SDSS optical and VLA radio images of SDSS J11129+6057. The {\it r}-band
image on the left shows the host galaxy, with its prominent bulge and faint disk. The
panel of radio maps on the right bring out the compact triple radio source, which is well resolved
only in the 22 GHz image. Note that the angular scale of the radio maps is 10 times smaller than the scale
in the SDSS image.
}
\end{figure*}

As part of the VLA program described in \S 2.2, we also imaged a second galaxy from sample of
\citet{smith09}: SDSS J112939.78+605742.6.
This object was detected in the NVSS survey with a 1.4 GHz flux density of $24.6 \pm 1$ mJy.
The SDSS spectrum of this galaxy shows clear double peaked [\ion{O}{3}] lines at a redshift of 0.112, a Type II AGN spectrum
and a relatively evolved stellar population, similar to \j1517, though the equivalent width of the strong emission lines are
several times lower than in \j1517. \citet{smith09} measure a [\ion{O}{3}]$\lambda 5007$ line-splitting of 550 \kms.
The SDSS images show a red host with a pronounced central bulge and relatively faint disk, 
seen at a mild inclination to the line of sight. No evidence can be seen in the images
for extended blue nuclear structures; the bulge isophotes appear circular and uniform in all SDSS images. 

In Fig.~\ref{appendixfig}, we plot the SDSS {\it r}-band image of J1129+6057 and radio maps in the four VLA bands.
The radio source is found to be quite compact and resolves into a linear triple structure in the 22 GHz map, reminiscent of
a classic core-jet geometry. The centroid of the galaxy bulge from the {\it r}-band image lies closest to the middle 
of the three radio knots and we identify that with a radio core. Unlike \j1517, the core in the J1129+6057 
radio source is not dominant and is actually weaker than the jets at 22 GHz. The overall size of the radio
source is $0\farcs403$ ($800$ pc) between the outer knots,  significantly more physically compact than the
radio jet in \j1517.

Since a clear extended jet structure is seen in this galaxy as well, we believe that the most likely
cause of the double-peaked narrow lines is due to a jet interaction. We find no signature of 
double black holes in this system: the triple structure of the radio source can be explained by an 
extended source geometry and the location of the core is essentially at the photometric center
of the host galaxy, while in a system of in-spiraling black holes, the core would most likely be
considerably offset from the galaxy's center. Unfortunately, the relative association between
emission line gas and the jets cannot be verified at this stage given the compactness 
of the radio source. Emission line mapping at 100 mas scale resolution is needed to do this.

The presence of extended jets in radio bright double-peaked narrow line AGN highlights the role that
jet interactions can play in producing disturbed emission line regions. The highly compact nature
of J1129+6057 shows that even sub-kpc radio jets can produce NLR-wide disturbances and
high resolution imaging is needed to fully study the region of influence of the jet.

\end{document}